\documentstyle[11pt,draft]{article}
\newcommand{\be}{\begin{equation}}
\newcommand{\ee}{\end{equation}}

\newcommand{\pro}[2]{\mbox{$\langle\, #1 , #2\,\rangle$}}

\newcommand{\real}{\mbox{{\rm I\hspace{-2truemm} R}}}

\title{Space--time singularities and the axion
in the Poincar\'e coset models ISO(2,1)/H}
\author{Roberto Casadio\thanks{
Permanent Institution: I.N.F.N., Sezione di Bologna, Italy,
e--mail: casadio@bologna.infn.it}
\ and Benjamin Harms\thanks{e--mail: bharms@ua1vm.ua.edu}\\
 \\
{\em
Department of Physics and Astronomy, The University of Alabama}\\
{\em Box 870324, Tuscaloosa, AL 35487-0324}}
\begin{document}
\baselineskip 4.0ex
\begin{titlepage}
\pagestyle{empty}
\maketitle
\begin{abstract}
By promoting an invariant subgroup $H$ of $ISO(2,1)$ to a gauge
symmetry of a WZWN action, we obtain the description of a bosonic
string moving either in a curved 4-dimensional space--time with
an axion field and curvature singularities or in 3-dimensional
Minkowski space--time.
\end{abstract}
Pacs: 04.20.DW, 11.25.-w, 11.30.Cp
\par\noindent
Keywords:\par
axion field\par
bosonic strings\par
coset construction\par
curved space--times\par
Poincar\'e invariance\par
signature changes\par
space--time singularities\par
spinning strings\par
\rightline{UAHEP 965}
\end{titlepage}
%
\pagestyle{plain}
\raggedbottom
\setcounter{page}{1}
\section{Introduction}
In recent years it has emerged that several string actions
naturally describe curved space--times with singularities.
This was first realized when Witten discovered that a gauged WZWN
action for $SL(2,\real)/U(1)$ contains a black hole \cite{witten}.
In a subsequent paper \cite{HH}, this model was extended
and found to be conjugated to a black string, and
more general coset models $G/H$, with $G$
a simple non--compact group, have been analyzed \cite{bars}.
\par
In this letter we apply the coset construction of
Refs.~\cite{witten,HH,bars} to a WZWN action in the Poincar\'e group
$ISO(2,1)$ that describes a closed bosonized spinning string in
2+1-dimensional Minkowski space--time \cite{stern} and
we argue that it leads to an effective theory with $6-{\rm dim}(H)$
degrees of freedom.
\par
We also show, in the framework of a particular parameterization,
that one further degree of freedom can be eliminated, giving the
description of a string without spin moving in a 4-dimensional
curved space--time with an axion field and curvature singularities or
in 3-dimensional Minkowski space--time.
\par
\setcounter{equation}{0}
\section{The gauged WZWN action}
The elements of the Poincar\'e group $ISO(2,1)$ can be written using
the notation $g=\left(\Lambda,v\right)$,
where $\Lambda\in SO(2,1)$ and $v\in \real^3$.
Given the map $g:\ M=D^2\times R\mapsto ISO(2,1)$ from the
2-dimensional disc$\times$time to $ISO(2,1)$,
we consider the WZWN action
\be
S={1\over 2\lambda^2}\,\int_{\partial M} d^2\sigma\,
\pro{g^{-1}\,dg}{g^{-1}\,dg}+
{n\over 12\,\pi}\,\int_M
\pro{g^{-1}\,dg}{(g^{-1}\,dg)^2}
\ ,
\ee
where $\partial M$ is the boundary of the manifold $M$, that is the
string world--sheet parameterized by the light--cone
coordinates $\sigma^+,\sigma^-$,
and the brackets $\pro{ }{ }$ denote the non--degenerate bilinear
invariant of $ISO(2,1)$.
If $-1/\lambda^2=n/4\,\pi\equiv\kappa/2$,
the action can be written entirely on
the boundary and it describes a closed bosonized spinning string
moving in 2+1 Minkowski space--time with coordinates
$v^k$, $k=0,1,2$ \cite{stern},
\begin{equation}
S=-{\kappa\over 4}\,\int_{\partial M} d^2\sigma\,
\epsilon^{ijk}\,
\left(\partial_+\Lambda\,\Lambda^{-1}\right)_{ij}\,
\partial_- v_k
\ ,
\label{S}
\end{equation}
where $\epsilon^{ijk}$ is the 3-dimensional Levi--Civita symbol.
\par
The action in Eq.~(\ref{S}) is invariant under
$g\mapsto h_{_L}(\sigma^+)\,g\,h_{_R}^{-1}(\sigma^-)$,
where $h_L$, $h_R\in ISO(2,1)$,
and also under the left and right action of the group of
diffeomorphisms of the world--sheet \cite{stern}.
However, it is not invariant under the local action
of any subgroup $H$ of $ISO(2,1)$ given by
$g\mapsto h_{_L}\,g\,h_{_R}^{-1}=
\left(\theta_{_L}\,\Lambda\,\theta^{-1}_{_R},
-\theta_{_L}\,\Lambda\,\theta^{-1}_{_R}\,y_{_R}+
\theta_{_L}\,v+y_{_L}\right)$,
where $h_{_{L/R}}=h_{_{L/R}}(\sigma^-,\sigma^+)=
(\theta_{_{L/R}},y_{_{L/R}})\in H$,
due to the dependence of $h_{_L}$ on $\sigma^-$
and of $h_{_R}$ on $\sigma^+$.
To promote $H$ to a gauge symmetry of the action we introduce
gauge fields $A_\pm=\left(\omega_\pm,\xi_\pm\right)\in iso(2,1)$,
the Lie algebra of the Poincar\'e group, and covariant derivatives
$D_\pm=\partial_\pm+A_\pm$.
\par
We also demand $H$ to be invariant, such that
$\delta g=h_{_L}\,g\,h_{_R}^{-1}\in H$.
The only possible choices for $ISO(2,1)$ are subgroups of
the translation group $\real^3$, that is
$h_{_{L/R}}=(0,y_{_{L/R}}^n)$, where $n$ runs in a subset
of $\{0,1,2\}$, for which $\delta g=(0,y_{_L})=h_{_L}\,g$.
In this case $\omega_\pm\equiv\xi_+ \equiv 0$, and
$\xi_-^k\equiv 0$ iff the translation in the $k$ direction is not
included in $H$.
The gauged action then reads
\be
S_g=-{\kappa\over 4}\,\int d^2\sigma\,\epsilon^{ijk}\,
\left(\partial_+\Lambda\,\Lambda^{-1}\right)_{ij}\,
\left(\partial_-v+\xi_-\right)_k
\ .
\label{S_g}
\ee
\par
For the ungaged action $S$ in Eq.~(\ref{S}) the equations of motion
$\delta_v S=0$, which follow from the variation $v\to v+\delta v$,
with $\delta v$ an infinitesimal 2+1 vector, lead to the
conservation of the momentum currents $P_+^k\equiv
\epsilon^{ijk}\,(\partial_+\Lambda\,\Lambda^{-1})_{ij}$,
$k=0,1,2$ \cite{stern}.
In the gauged case this variation must be supplemented
by the condition that the gauge field varies under an
infinitesimal $ISO(2,1)$ transformation according to
\be
\xi_-^n\to \xi_-^n-\partial_-(\delta v^n)
\ ,
\label{cond}
\ee
and from $\delta_v S_g=0$ one obtains
\be
\partial_-P_+^{k\not= n}=0
\ ,
\label{dP}
\ee
so that only the $P_+^{k\not= n}$ currents are still conserved.
\par
Similarly, from the variation $\Lambda\to\Lambda+\delta\Lambda$,
$\delta\Lambda=\Lambda\,\epsilon$ and $\delta v=\epsilon\,v$,
with $\epsilon_{ij}=-\epsilon_{ji}$ an infinitesimal $so(2,1)$
matrix, the equations $\delta_\epsilon S=0$ lead to the conservation
of the three angular momentum currents
$J_-^k\equiv\left(\Lambda\,\partial_-v\right)^k$, $k=0,1,2$,
which can be shown to include a contribution of intrinsic
(non orbital) spin \cite{stern}.
In the gauged case, by making use of Eq.~(\ref{cond}) and
Eq.~(\ref{dP}), one obtains
\be
\partial_+J_-^k=-\partial_+(\Lambda\,\xi_-)^k
\ ,
\label{dJ}
\ee
so that the currents $J_-^k$ couple to the gauge field.
\par
We are free to choose ${\rm dim}(H)$ gauge conditions to be
satisfied by the elements of $ISO(2,1)/H$.
It appears natural to impose
\be
\xi^n_-=-\partial_- v^n
\label{gauge}
\ ,
\ee
so that the previous equations of motion become the same as
$\delta_v S_{eff}=\delta_\epsilon S_{eff}=0$ obtained by varying
the effective action
\begin{equation}
S_{eff}=-{\kappa\over 2}\,\int d^2\sigma\,
\sum\limits_{k\not=n} P_+^k\,\partial_-v_k
\ ,
\label{S_eff}
\end{equation}
where the sum runs only over the indices corresponding to the
translations not included in $H$.
\par
We observe that, although the number of degrees of freedom in the
effective action is ${\rm dim}(ISO(2,1))-{\rm dim}(H)=6-{\rm dim}(H)$
if ${\rm dim}(H)<3$,
gauging the whole 3-dimensional translation subgroup makes the
effective theory empty and the present reduction scheme fails.
Thus we shall attempt at gauging 1-and 2-dimensional translations
only.
\subsection{Gauging a 1-dimensional translation}
In order to obtain an explicit form for the momentum currents
we write an $SO(2,1)$ matrix $\Lambda^i_{\ j}$
as a product of two rotations (of angles $\alpha$ and $\gamma$)
and a boost ($\beta$) \cite{hrt},
\begin{eqnarray}
\Lambda&=&
\left[\begin{array}{ccc}
1 & 0 & 0 \\
0 & \cos\alpha & -\sin\alpha \\
0 & \sin\alpha & \cos\alpha
\end{array}\right]\,
\left[\begin{array}{ccc}
\cosh\beta & 0 & \sinh\beta \\
0 & 1 & 0 \\
\sinh\beta & 0 & \cosh\beta
\end{array}\right]\,
\left[\begin{array}{ccc}
1 & 0 & 0 \\
0 & \cos\gamma & -\sin\gamma \\
0 & \sin\gamma & \cos\gamma
\end{array}\right]
\nonumber \\
&=&
\left[\begin{array}{ccc}
\cosh\beta & \sinh\beta\,\sin\gamma & \sinh\beta\,\cos\gamma \\
-\sin\alpha\,\sinh\beta
&\cos\alpha\,\cos\gamma-\sin\alpha\,\cosh\beta\,\sin\gamma
&-\cos\alpha\,\sinh\beta-\sin\alpha\,\cosh\beta\,\sin\gamma \\
\cos\alpha\,\sinh\beta
&\sin\alpha\,\cos\gamma+\cos\alpha\,\cosh\beta\,\sin\gamma
&-\sin\alpha\,\sin\gamma+\cos\alpha\,\cosh\beta\,\cos\gamma
\end{array}\right]
\ ,
\end{eqnarray}
and we obtain
\be
\begin{array}{l}
P^0_+=
-\partial_+\alpha-\cosh\beta\,\partial_+\gamma \\
 \\
P^1_+=
-\cos\alpha\,\partial_+ \beta
-\sin\alpha\,\sinh\beta\,\partial_+\gamma  \\
 \\
P^2_+=
-\sin\alpha\,\partial_+\beta+\cos\alpha\,\sinh\beta\,\partial_+\gamma
\ .
\end{array}
\label{P}
\ee
\par
Then we choose $H=\{(0,y^0)\}$ and, since no derivative of $\alpha$
occurs in $P_+^1$ and $P_+^2$, we also rotate the variables $v^1$ and
$v^2$ by an angle $-\alpha$,
\be
\left[\begin{array}{c}
\partial_-\tilde v^1 \\
\partial_-\tilde v^2
\end{array}\right]
\equiv\left[\begin{array}{cc}
\cos\alpha & \sin\alpha \\
-\sin\alpha & \cos\alpha
\end{array}\right]\,
\left[\begin{array}{c}
\partial_-v^1 \\
\partial_-v^2
\end{array}\right]
\ .
\ee
This can be considered as an internal symmetry of the effective
theory which is used to further simplify the effective action in
Eq.~(\ref{S_eff}) with $n=0$ to the form
\begin{equation}
S_{eff}=-{\kappa\over 2}\,\int d^2\sigma\,\left[
-\partial_+\beta\,\partial_-\tilde v^1+
\sinh\beta\,\partial_+\gamma\,\partial_-\tilde v^2\right]
\ .
\label{S_4}
\end{equation}
Now we can introduce new string coordinates $x^i$, $i=1,\ldots,4$,
defined by
\be
\left\{\begin{array}{l}
4\,x^1=\beta+\gamma+v^1+v^2\\
4\,x^2=\beta-\gamma+v^1+v^2\\
4\,x^3=\beta+\gamma-v^1+v^2\\
4\,x^4=\beta+\gamma+v^1-v^2
\ ,
\end{array}\right.
\ee
and the action in Eq.~(\ref{S_4}) finally becomes
\begin{equation}
S_{eff}=-{\kappa\over 4}\,\int d^2\sigma\,\left[
G_{ij}\,\partial_+x^i\,\partial_-x^j+
B_{ij}\,(\partial_+x^i\,\partial_-x^j-
\partial_-x^i\,\partial_+x^j)\right]
\ ,
\label{S_4f}
\end{equation}
where ($f(\beta)\equiv\sinh\beta$)
\be
G_{ij}(\beta)=\left[\begin{array}{cccc}
2\,\left(f-1\right) & -1 & 0 & -1 \\
-1 & -2\,\left(f+1\right) & 0 & \left(f-1\right) \\
0 & 0 & 2\,\left(f+1\right) & 0 \\
-1 & \left(f-1\right) & 0 & -2\,\left(f+1\right)
\end{array}\right]
\ .
\ee
and it can be regarded as the metric tensor of our 4-dimensional
space--time.
The matrix $B_{ij}=B_{ij}(\beta)$ is antisymmetric
and defined by
\be
\begin{array}{lcl}
B_{12}=f
&\ \ \ \ \ \ &
B_{23}=1-f \\
B_{13}=1
&\ \ \ \ \ \ &
B_{24}=0 \\
B_{14}=-f
&\ \ \ \ \ \ &
B_{34}=-(f+1)
\ .
\end{array}
\ee
It represents an axion field whose field strength,
$H_{ijk}\equiv\partial_i B_{jk}+\partial_j B_{ki}+
\partial_k B_{ij}$, has $H_{124}=2\,\cosh\beta$ as
the only non--zero component.
\par
One can also show that the Ricci tensor $R_{ij}$ is not zero and
the curvature scalar,
\be
R(\beta)={27\,f^7+186\,f^6+49\,f^5-179\,f^4+87\,f^3+16\,f^2-17\,f-3
\over (3\,f+1)^2\,(f+1)^2\,(f^2+2\,f-2)^2}
\ ,
\ee
has singularities in $f\equiv\sinh\beta=-1,-1/3,-1\pm\sqrt{3}$
(see Fig.~1 for a plot of $R(\beta)$).
Thus ${\bf G}$ is not a vacuum solution of the gravitational
equations, and the action in Eq.~(\ref{S_4f}) is no longer conformally
invariant.
A dilaton field $\Phi$ must be introduced according to the general
1-loop expression $R_{ij}=\nabla_i\nabla_j\Phi$, plus the central
term $(d-26)/3=22/3$ due to the dimension $d=4$ of space--time
\cite{witten,gsw}.
\par
The signature of the metric is $3+1$ in a neighborhood of
$\beta=0$, but it changes in coincidence with the curvature
singularities, as can be inferred by noting that the determinant,
\be
G\equiv\det({\bf G})=4\,(3\,f+1)\,(f+1)\,(f^2+2\,f-2)
\ ,
\ee
is proportional to the square root of the denominator of $R$
(see Fig.~2 for a plot of $G(\beta)$).
\subsection{Gauging 2-dimensional translations}
We choose to gauge $H=\{(0,y^1),(0,y^2)\}$.
Since no derivative of $\beta$ occurs in $P_+^0$,
by defining $\partial_-w\equiv\cosh\beta\,\partial_-\gamma$,
we can eliminate it.
Introducing $3\,x^1\equiv \alpha+w+v^0$, $3\,x^2\equiv \alpha-w+v^0$
and $3\,x^3\equiv \alpha+w+v^0$ the effective action in
Eq.~(\ref{S_eff}) with $k=0$ can be written in the same form as
Eq.~(\ref{S_4f}),
but now $G_{ij}$ is a constant symmetric matrix with signature 2+1
and $B_{ij}$ is a constant antisymmetric matrix.
\par
Again, if we interprete $x^1,x^2,x^3$ as the coordinates of
a string, we conclude that the target space is the same
2+1 Minkowski space--time of the ungaged action in Eq.~(\ref{S}),
but the string has lost its intrinsic spin and the axion field
is a pure gauge, $H_{ijk}\equiv 0$.
\section{Conclusions}
One of the main aspects of the model we have studied is that its
effective action naturally contains an axion field together with a
bosonic string in (possibly) curved backgrounds.
\par
In the 3-dimensional case that we have studied, the outcome of the
coset construction seems to be quite trivial, leading solely to the
initial flat space--time with the string now deprived of its
intrinsic spin.
However the 4-dimensional case shows regions of different signatures
(including Minkowskian regions) separated by curvature
singularities.
\par
Further, by gauging different translations one can obtain
(5- and 4-dimensional) solutions other than the ones we have shown
here.
We are at present trying to perform a complete analysis of the
model, including the use of different parameterizations of the
Lorentz group.
\par
\bigskip
\par
\centerline{\large Acknowledgements}
\smallskip
\par
This work was supported in part by the Department of Energy
under contract n.~DE-FG02-96ER40967.
We would like to thank A. Stern for many useful discussions.
\end{document}